\documentclass{jpp}

\usepackage{graphicx}
\usepackage{natbib}
\usepackage{rotating}
\usepackage{amsmath}
\usepackage{float}
\usepackage{hyperref}

\ifCUPmtlplainloaded \else
  \checkfont{eurm10}
  \iffontfound
    \IfFileExists{upmath.sty}
      {\typeout{^^JFound AMS Euler Roman fonts on the system,
                   using the 'upmath' package.^^J}%
       \usepackage{upmath}}
      {\typeout{^^JFound AMS Euler Roman fonts on the system, but you
                   dont seem to have the}%
       \typeout{'upmath' package installed. JPP.cls can take advantage
                 of these fonts, if you use 'upmath' package.^^J}%
      }
  \else
  \fi
\fi

\ifCUPmtlplainloaded \else
  \checkfont{msam10}
  \iffontfound
    \IfFileExists{amssymb.sty}
      {\typeout{^^JFound AMS Symbol fonts on the system, using the
                'amssymb' package.^^J}%
       \usepackage{amssymb}%

      }{}
  \fi
\fi

\ifCUPmtlplainloaded \else
  \IfFileExists{amsbsy.sty}
    {\typeout{^^JFound the 'amsbsy' package on the system, using it.^^J}%
     \usepackage{amsbsy}}
    {\providecommand\boldsymbol[1]{\mbox{\boldmath $##1$}}}
\fi

\newcommand{\farcm}{\mbox{\ensuremath{.\mkern-4mu^\prime}}}

\newsavebox{\astrutbox}
\sbox{\astrutbox}{\rule[-5pt]{0pt}{20pt}}

\newcommand{\bc}{\begin{center}}
\newcommand{\ec}{\end{center}}
\newcommand{\be}{\begin{equation}}
\newcommand{\ee}{\end{equation}}
\newcommand{\ba}{\begin{eqnarray}}
\newcommand{\ea}{\end{eqnarray}}
\newcommand{\bt}{\begin{tabular}}
\newcommand{\et}{\end{tabular}}

\newcommand{\chan}{{\sl Chandra}}

\def\farcm{\hbox{$.\!\!^{\prime}$}}

\def\chan{{\sl Chandra}}

\title[Pulsar wind nebulae 
created by fast-moving pulsars]{Pulsar wind nebulae 
 created by fast-moving pulsars}

\author[O.\ Kargaltsev, G.\ G.\ Pavlov,  N.\ Klingler, and B.\ Rangelov]%
{O.\ns K\ls A\ls R\ls G\ls A\ls L\ls T\ls S\ls E\ls V$^1$%
  \thanks{Email address for correspondence: kargaltsev@gwu.edu},\ns
G.\ns G.\ns P\ls A\ls V\ls L\ls O\ls V$^2$,\break
N.\ns  K\ls L\ls I\ls N\ls G\ls L\ls E\ls R$^1$,
\and\ns  B.\ns R\ls A\ls N\ls G\ls E\ls L\ls O\ls V$^{1,3}$}

\affiliation{$^1$Department of Physics, George Washington University,
Washington, DC 20052, USA\\[\affilskip]
$^2$Department of Astronomy \& Astrophysics, Pennsylvania State University, University Park, PA 16802, USA\\[\affilskip]
$^3$Department of Physics, Texas State University, San Marcos,  TX 78666, USA}

\pubyear{2010}
\volume{650}
\pagerange{119--126}
\date{?; revised ?; accepted ?. - To be entered by editorial office}
\begin{document}

\maketitle

\begin{abstract}
We review multiwavelength properties of pulsar wind nebulae (PWNe) created by super-sonically moving pulsars and the effects of pulsar motion on the PWN morphologies and the ambient medium.
Supersonic pulsar wind nebulae (SPWNe) are characterized by bow-shaped shocks around the pulsar and/or cometary tails filled with the shocked pulsar wind. 
In the past several years significant advances in SPWN studies have been made in deep observations with the {\sl Chandra} and {\sl XMM-Newton} X-ray Observatories as well as the {\sl Hubble Space Telescope}.  
In particular, these observations have revealed very diverse SPWN morphologies in the pulsar vicinity, different spectral behaviors of long pulsar tails, the presence of puzzling outflows misaligned with the pulsar velocity, and far-UV bow shocks. 
Here we review the current observational status focusing on recent developments and their implications. 
\end{abstract}


\section{Introduction}
Pulsars and their winds are among nature's most powerful particle accelerators, producing particles with energies up to a few PeV.  
 In addition to high-energy radiation from the magnetosphere, the 
 rotational (spin) energy of a neutron star (NS) is carried away in the form of a magnetized ultra-relativistic pulsar wind (PW), whose nonthermal emission can be seen from radio to $\gamma$-rays, with energies reaching nearly 100 TeV \citet{Aharonian04}.
Recent X-ray and $\gamma$-ray observations suggest that the ratio of 
 energy  
  radiated from the ``observable'' PWN 
  to the energy radiated from the magnetosphere can vary significantly among pulsars. 
 For instance, for the Crab pulsar and its PWN the ratio of the luminosities integrated over the entire electromagnetic spectrum\footnote{The estimates of the Crab's PWN and pulsar luminosities are based on the spectra shown in Figure 2 of  \citet{2014RPPh...77f6901B}. The  Vela PWN luminosity is estimated from the spectrum shown in Figure 4  of \citet{Mattana11} which corresponds to the $r=6'$ region around the pulsar. The Vela pulsar's luminosity is based on Figure 28 of \citet{2015MNRAS.449.3827K}.} is $ L_{\rm PWN}/L_{\rm PSR}\simeq30$  while for the Vela pulsar and its compact PWN it is   $\simeq0.03$. The ratio becomes $\sim 1$ if the luminosity\footnote{ Vela X is believed to be a relic PWN filled with  PW particles which were produced when the pulsar had a higher $\dot{E}$. The  Vela X  luminosity is estimated from the spectrum shown in Figure 4  of \citet{Mattana11}}  of  Vela X (a large structure south of the pulsar bright in radio, X-rays, and  high energy $\gamma$-rays; \citealt{2013ApJ...774..110G})  is added to the compact PWN luminosity.

 The   speed of the PW
 is 
 highly relativistic immediately beyond the pulsar magnetosphere.
 However, the  interaction  with the 
 ambient medium  causes the wind to slow down abruptly at a termination shock (TS). 
Immediately downstream of the TS, the flow speed is expected to
 become mildly
relativistic,
  lower than the speed of sound in ultra-relativistic magnetized  plasma but 
 much higher than the   sound speed in the ambient medium. 
 It is commonly assumed that the distance to the TS from the pulsar, $R_{\rm TS} \sim ( \dot{E}/4\pi c P_{\rm amb}  )^{1/2}$, can be estimated by balancing the PW pressure\footnote{Although this is a commonly used estimate, rigorously speaking, instead of $\dot{E}$ one should use a fraction of $\dot{E}$ which is associated with the wind and can vary from pulsar to pulsar (see the discussion above). } with the ambient pressure. 
  The flow keeps decelerating further downstream 
and reaches the contact discontinuity (CD) 
 that separates the shocked PW from the  surrounding medium shocked in the forward shock (FS)\footnote{In reality, various instabilities may lead to a more complex picture where the CD is distorted and possibly destroyed.}.
   For a stationary (or slowly moving, subsonic) pulsar, 
 the CD sphere (hence the PWN)  is 
  expected to be 
 expanding as the pulsar pumps more energy 
into it, until the radiative and/or adiabatic expansion losses  balance the energy input.
 The shocked PW
 within the PWN bubble contains ultra-relativistic particles with randomized pitch angles and a magnetic field whose structure also becomes somewhat disordered (see 3D simulations of the Crab PWN; \citealt{Porth16}).  
Therefore,  PWNe are expected to produce 
 synchrotron radiation, responsible for the PWN emission from
 radio frequencies through X-rays and into the MeV range,
  and inverse-Compton (IC) radiation in the GeV-TeV range
 (see e.g., the reviews by \citealt{KP08}, \citealt{KP10}, \citealt{Kargaltsev15}, and \citealt{Reynolds17}).

In general, studying PWNe provides 
 information about the pulsars that power them, the properties of the surrounding 
 medium, and the physics of the wind-medium interactions. 
The detailed structure of the interface between the PW
 and the surrounding ambient medium 
  is not well understood. 
In an idealized hydrodynamic scenario, 
 the CD separates the shocked ISM 
  from the shocked PW.
 Being compressed and heated at the FS, 
 the shocked ambient medium emits   
 radiation in spectral lines and continuum. 
 It has been notoriously difficult to identify the FS around many young and bright PWNe residing in SNRs (including the Crab PWN).
 However, there is a class of PWNe, 
 associated with supersonically moving pulsars, where both the 
 shocked ambient medium and the 
 shocked PW can be seen. 
In this review we will focus on observational properties of supersonic pulsar wind nebulae (SPWNe). 
For a recent theoretical review, see \citet{Bykov17}.

\subsection{SPWNe -- PWNe of supersonic pulsars}
 In addition to the ISM pressure, the PWN size and morphology can be significantly affected by the ram pressure of the external medium caused by the fast (supersonic) pulsar motion. 
 Indeed, average pulsar 3D velocities have been found to be $v_p\sim400$
 km s$^{-1}$  
 for an isotropic velocity distribution \citep{Hobbs05}. 
This implies that the majority of pulsars only stay within their host SNR environment for a few tens of 
 kilo-years, although some particularly fast-moving pulsars can leave it even earlier.
 Once the pulsar leaves its host SNR, it 
 enters a very different environment,
 with a much lower sound speed, 
$c_s\sim 3$--30 km s$^{-1}$, depending on the ISM phase\footnote{The sound speed in a middle-aged SNR can be on the order of a few hundred km s$^{-1}$.}, 
  hence the pulsar motion becomes highly supersonic,
 i.e., $v_p/c_s\equiv \mathcal{M}
 \gg 1$, where $\mathcal{M}$ is the Mach number.

The supersonic motion strongly modifies the PWN appearance and the
properties of its emission, making it useful to introduce a separate category of supersonic pulsar wind nebulae (hereafter SPWNe).  
 In particular, for an isotropic  PW, 
 the FS, CD, and TS
shapes resemble paraboloids in the pulsar vicinity but have quite
different shapes behind the pulsar (see Figure 9 in \citealt{Gaensler04}). 
The distance from the pulsar to the apex of the CD can be estimated as 
\be R_a \approx \left[\frac{\dot{E} f_\Omega}{4\pi c (P_{\rm amb}+P_{\rm ram})}\right]^{1/2}. 
\ee
  At this distance, the pulsar wind pressure, $P_{w} = \dot{E} f_\Omega (4\pi c r^2)^{-1}$
  ($f_\Omega$ takes into account PW anisotropy),
 is balanced by the sum of the ambient pressure, $P_{\rm amb}= \rho kT (\mu m_{\rm H})^{-1}= 1.38\times 10^{-12} n_{\rm H} \mu^{-1} T_4$ 
 dyn cm$^{-2}$, and the ram pressure, $P_{\rm ram}= \rho v^2 = 1.67\times 10^{-10} n v_7^2$ 
 dyn cm$^{-2}$ ($T_4 = T/10^4\,{\rm K}$, $v_7 = v/10^7\,{\rm cm\, s}^{-1}$, $\mu$ is the mean molecular weight, and $n = \rho/m_{\rm H}$ is in units of cm$^{-3}$). 
Assuming $P_{\rm ram}\gg P_{\rm amb}$ (or ${\mathcal M} \gg 1$), 
 we obtain $R_{a}= 6.5\times 10^{16}n^{-1/2} f_\Omega^{1/2} \dot{E}_{36}^{1/2} v_7^{-1}$ cm (see e.g., \citealt{KP07}).

The shocked pulsar wind, whose synchrotron emission 
 can be seen in X-rays and radio, is confined between the TS and CD surfaces. 
For ${\mathcal M}\gg 1$ and a {\em nearly isotropic} pre-shock wind with a small magnetization parameter 
 (see \S2.2), the TS acquires a bullet-like shape (\citealt{Gaensler04,Bucciantini05}).
 The length of the 
 bullet and the 
 diameter of the post-TS PWN
  are $\simeq(5-6)R_a$ and 
    $\simeq 4 R_a$, respectively 
    \citep{Bucciantini05}.

Additional complexity arises due to the fact that the PW
 is 
 likely not isotropic but concentrated towards the
equatorial plane of the rotating pulsar. Evidence of such anisotropy,
at least in young pulsars, is seen in high-resolution
 \chan\ images (e.g., \citealt{KP08}), 
and it is 
 supported by theoretical modeling \citep{Komissarov04}. 
Therefore the appearance of an  
SPWN  
is expected to depend on the angle between the velocity vector and the spin axis of the pulsar.  
\citet{Vigelius07} demonstrated this by performing 3D hydrodynamical
 simulations and introducing a latitudinal dependence of the wind power (with the functional form expected for  the split vacuum dipole  solution). 
In addition, inhomogeneities in the ambient medium \citep{Vigelius07} and ISM entrainment \citep{Morlino15} are expected to affect the bow shock shape. 
Observational confirmation of these  
 expectations comes from X-ray and H$\alpha$ (and more recently also far-UV, see below) images of SPWNe.

\begin{figure}
\centerline{\includegraphics[scale=0.73]{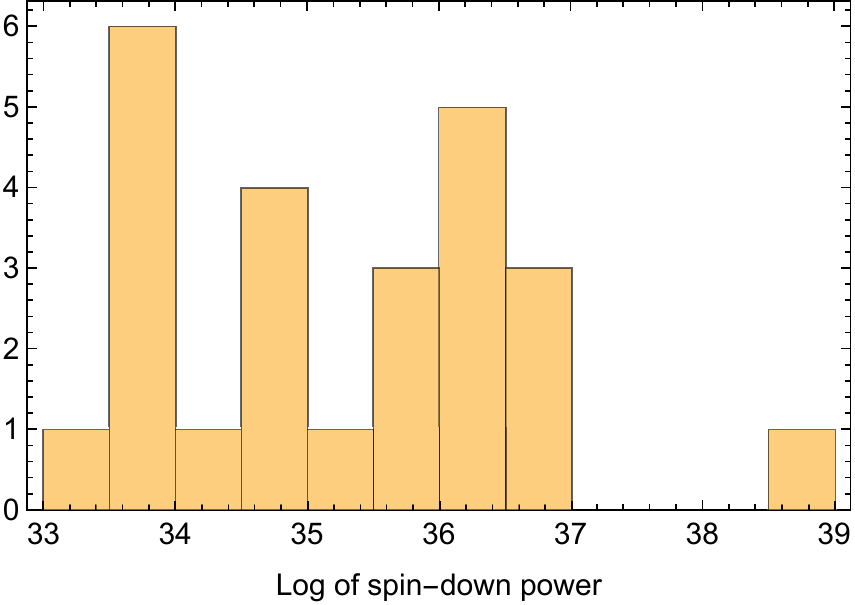}
\includegraphics[scale=0.73]{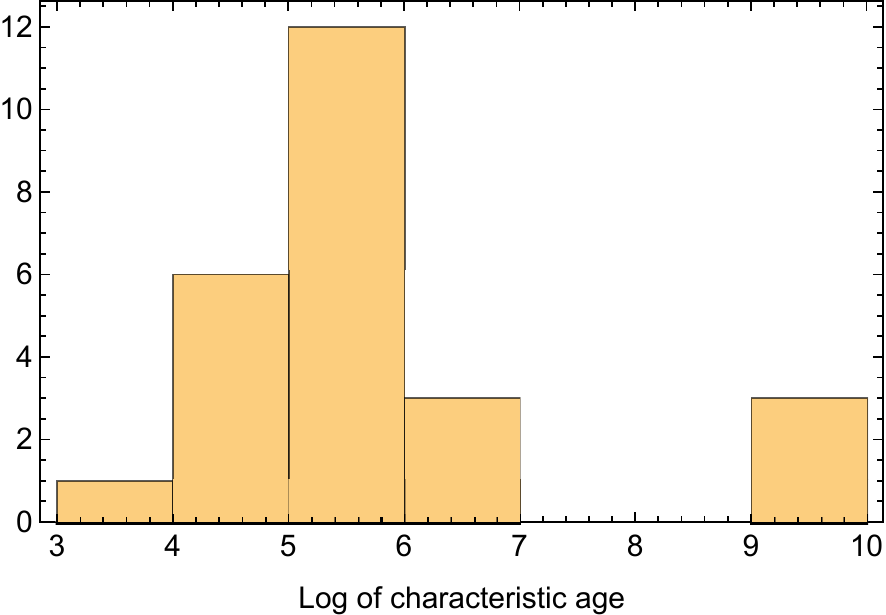}}
\caption{$\dot{E}$ (left; in units of erg s$^{-1}$) and age (right; in years) distributions for the pulsars producing the SPWNe from Table 1.}
 \label{fig:histo}
\end{figure}

\section{Current Sample of SPWNe}

\begin{figure}
\vspace{-1cm}
\centerline{\includegraphics[scale=0.46]{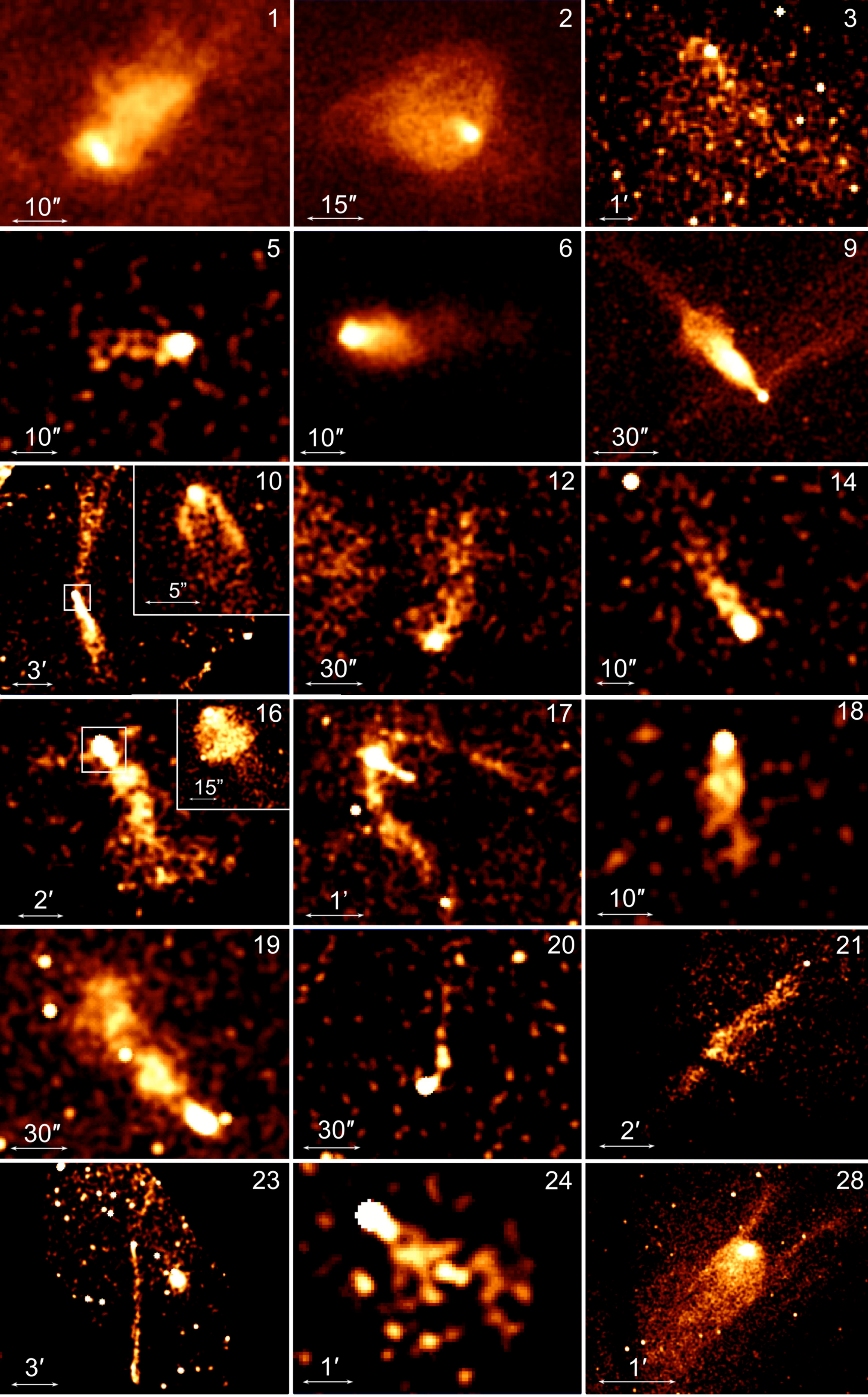}}
\caption{{\sl Chandra} ACIS images of 18 SPWNe.  The panels are numbered in accordance with Tables 1 and 2.  {\sl Chandra} images of some of these  objects are also shown in \citet{Reynolds17}.}
 \label{fig:bow:xray}
\end{figure}

Currently, there are about 30 pulsars whose  X-ray, radio, or $H\alpha$ images either clearly show or strongly suggest effects of supersonic motion (see Tables 1 and 2).
 Most of these SPWNe (or SPWN candidates) have been found in X-rays, primarily through high-resolution imaging with {\sl Chandra} (see Figure \ref{fig:bow:xray}).
In 8 cases the  supersonic  PWN morphologies can be seen in radio. In 2 cases
(PSRs  B0906--49 and J1437--5959)  radio images clearly show SPWNe 
 which are not seen in  X-ray images.   
 Finally, there are  8 rotation-powered pulsars with detected H$\alpha$ bow shocks, of which 7 have X-ray PWN detections (see Table 2), with two 
   bow shocks also detected in far-UV.   
Despite the existence of TeV observations, no SPWN detection in TeV has been reported yet.
With the exception of the very young and energetic PSR J0537--6910 in the LMC ($\tau=4.9$ kyr, $\dot{E}=4.9\times10^{38}$ erg s$^{-1}$), which may turn out to be not an SPWN despite the suggestive X-ray PWN morphology\footnote{The remote LMC location of the pulsar makes it difficult to reliably determine the pulsar velocity (from the PWN morphology or proper motion) and the properties of the ambient medium.}, the rest of SPWNe are powered by  non-recycled pulsars with ages between 10 kyr and 3 Myr, or by much older recycled pulsars; their spin-down powers span the range of $10^{33}$--$10^{37}$ erg s$^{-1}$  (see Figure \ref{fig:histo}).

\begin{table}
\begin{center}
\def~{\hphantom{0}}
\begin{tabular}{cllccccc}
\# & Pulsar & Associated Object(s) & $d$ & $\log\dot{E}$ & log $\tau$ & $B_{11}$ & $v_{\perp}$  \\ 
&  &              &   kpc    &      erg s$^{-1}$ &   yrs   &  $10^{11}$~G &  km~s$^{-1}$ \\
\hline
1  &    J0537--6910$^{\rm a}$ & SNR N157B  &  49.7  &  38.68  &  3.69  &  9.25 &    ...   \\
2  &    B1951+32  & SNR CTB 80  &  3  &  36.57  &  5.03  &  4.86 &    460 \\
3  &    J1826--1256 & HESS J1825--137  & $\sim$ 3.9$^{\rm b}$  &  36.56  &  4.16  &  37  &    ...  \\
4  &    B1706-44  & SNR G343.1--2.3 &  2.6  &  36.53  &  4.24  &  31.2  &    $\lesssim100$    \\
5  &    B1757--24  & SNR G5.27--0.9, Duck PWN  &  3.8  &  36.41  &  4.19  &  40.4 &   198  \\
6  &    J1747--2958  & Mouse PWN  &  5  &  36.40  &  4.41  &  24.9 &    $306\pm43$ \\
7  &    J1135--6055  &  ... &    $\sim$2.8$^{\rm b}$    &  36.32  &  4.36  &  30.5 &    $<330$ \\
8  &    J1437--5959 & SNR G315.9--0.0, Frying Pan PWN   &  8  &  36.15  &  5.06  &  7.37 &    $\sim 300$ \\
9  &    J1101--6101 & Lighthouse Nebula, SNR G290.1--0.8   &   $\sim$7$^{\rm b}$  &  36.13  &  5.06  &  7.24 &   $\sim$2000 \\
10  &    J1509--5850 & ...  &  4  &  35.71  &  5.19  &  9.14  &    $200-600$ \\
11  &    B0906--49  & ... &  1  &  35.69  &  5.05  &  12.9 &    $\sim60$ \\
12  &    B1853+01$^{\rm a}$  & SNR W44   &  3.3  &  35.63  &  4.31  &  75.5  &    $400^{+114}_{-73}$  \\
13  &    B0740--28 &  ... &  2  &  35.28  &  5.2  &  16.9 &  275$^{\rm c}$ \\
14  &    B1957+20 & the Black Widow pulsar &  1.73  &  35.20  &  9.18  &  0.002 &    $\sim220$  \\
15  &    J0538+2817 & SNR S147  &    $1.39^{\rm p}$    &  34.69  &  5.79  &  7.33  &    $357^{+59}_{-43}$ \\
16  &    B0355+54  & Mushroom PWN  &    $1.04^{\rm p}$    &  34.66  &  5.75  &  8.39  &     $61^{+12}_{-9}$  \\
17  &    J0633+1746  & Geminga PWN  &    $0.25^{\rm p}$    &  34.51  &  5.53  &  16.3 &     $\sim$200  \\
18  &    J2030+4415 &  ... &    $\sim1$$^{\rm b}$    &  34.46  &  5.74  &  12.3  &    ...    \\
19  &    J1741--2054 &  ... &  0.3  &  33.97  &  5.59  &  26.8   &     155 \\
20  &    J2124--3358  & ... &  0.41  &  33.83  &  9.58  &  0.003  &  75$^{\rm c}$\\
21  &    J0357+3205  & Morla PWN &  0.5  &  33.77  &  5.73  &  24.3 &    $\sim$2000\\
22  &    J0437--4715 &  ... &    $0.156^{\rm p}$    &  33.74  &  9.2  &  0.006  &    $104.7\pm0.9$\\
23  &    J2055+2539$^{\rm a}$  & ... &    $\sim0.6$$^{\rm b}$    &  33.69  &  6.09  &  11.6  &    $\lesssim2300$ \\
24  &    B1929+10 & ...  &    $0.36^{\rm p}$    &  33.59  &  6.49  &  5.18 &    $177^{+4}_{-5}$ \\
25  &    B2224+65 & Guitar Nebula &  1.88  &  33.07  &  6.05  &  26 &   1626 \\
26  &  ...  & SNR IC443$^{\rm a}$    &  1.4  &    ...    &   ...   &    ...   &    $\sim250$\\
27  &  ...  & SNR MSH 15--56, G326.3--1.8  &  4  &    ...    &   ...   &    ...  &     $100$--$400$ \\
28  &  ...  & G327.1--1.1, Snail PWN    &  7  &    ...    &   ...   &    ... &     $\sim500$  \\
\hline
\end{tabular}
\caption{Parameters of pulsars with SPWNe (from the ATNF catalog; \citealt{Manchester05}).
 The pulsars are listed in order of decreasing $\dot{E}$.
Distance $d$ is given in units of kpc, spin-down energy loss rate $\dot{E}$, pulsar characteristic age $\tau=P/2\dot{P}$,  surface magnetic field $B_{11}$, and projected pulsar velocity $v_\perp$.  {\bf Note:} Unless specified otherwise the pulsar distances are inferred from the dispersion measure (DM) according to \citet{2017ApJ...835...29Y} or taken from the individual papers (see references in Table 2). $^{\rm a}$ The supersonic nature of the PWNe powered by these pulsars has not been firmly established. $^{\rm p}$ Distances are from parallax measurements.  $^{\rm b}$ These are radio quiet pulsars detected in $\gamma$-rays with particularly uncertain distances.  $^{\rm c}$  Velocities are based on the estimates from \citet{Brownsberger14}. 
 }
\label{tab:kd}
\end{center}
\end{table}

\begin{table}
\begin{center}
 \def~{\hphantom{0}}
\begin{tabular}{clcccccccc}
 \# & Pulsar & $r_{\rm BS}$ &  $i$ & $l$ & $\log L_X$ & $\log\eta_{\rm X}$ & H$\alpha$ & Rad. & Ref. \\
& & 10$^{16}$ cm   & deg    &  pc  &  erg~s$^{-1}$  &   &  &   &\\
 \hline
1  &    J0537--6910$^{\rm b}$    &    ...      &    ...    &  3.7  &    $36.21\pm0.01$    &    $-2.47$    &    N    &    N    &    [1] \\  
2  &    B1951+32    &    $<12$        &    ...    &  1.2  &    $33.02\pm0.11$    &    $-3.55$    &    ?$^{\rm a}$    &    Y$^{\rm a}$    &    [2,3] \\ 
3  &    J1826--1256    &    ...    &    ...      &  5.8  &    $33.38\pm0.06$    &    $-3.18$    &    ?    &    ?    &    [52,53]\\   
4  &    B1706-44    &    $\sim70$     &    ?    &  3  &    $32.60\pm0.10$    &    $-3.93$    &    N    &    Y    &    [4] \\ 
5  &    B1757--24    &    $<13$    &    ...       &  0.4  &    $33.20\pm0.14$    &    $-3.21$   &    ?    &    Y    &    [3,35,36]  \\  
6  &    J1747--2958    &    $\sim14$        &    $\sim20^\circ$    &  1.1  &    $33.83\pm0.09$    &    $-2.57$    &    ?    &    Y    &    [5-8]  \\   
7  &    J1135--6055    &    ...        &    ...    &  1.4  &    $32.40\pm0.04$    &    $-3.92$    &    ?    &    ?    &    [53,54]\\  
8  &    J1437--5959    &    ...        &    ...   &    $\sim20$    &    ...    &    ...    &    ?    &    Y    &    [27] \\  
9  &    J1101--6101    &  34      &    ...    &  3.5  &    $32.40\pm0.40$    &    $-3.31$    &    ?    &    Y    &    [37-40]  \\ 
10  &    J1509--5850    &    $<0.6$       &    $\sim90^\circ$     &  6.5  &    $33.05\pm0.04$    &    $-2.66$    &    Y    &    Y    &    [9-11] \\  
11  &    B0906--49    &    ...        &    $\sim90^\circ$    &  3.5  &    ...     &   $-5.86$    &    ?    &    Y    &    [50,51]\\  
12  &    B1853+01$^{\rm b}$    &    $\sim33$      &   ...    &  1.3  &    $32.20\pm0.10$    &    $-2.58$    &    N    &    N    &    [12,13] \\      
13  &    B0740--28    &    $<6$    &    ...    &  ...  &    ...    &    ...    &    Y    &    ?    &    [26] \\ 
14  &    B1957+20    &  1.3      &    ...    &  0.3  &    $29.73\pm0.40$    &    $-5.14$    &    Y    &    ?    &    [33,34] \\   
15  &    J0538+2817    &    $<14$        &    $\sim90^\circ$    &  0.02  &    $31.30\pm0.15$    &    $-3.39$    &    N    &    N    &    [14,15]  \\  
16  &    B0355+54    &  0.5     &    $< 20^\circ$    &  1.5  &    $31.20\pm0.07$    &    $-3.46$    &    ?    &    N    &    [16,17]  \\  
17  &    J0633+1746    &  4     &    $>50^\circ$    &  0.35  &    $29.35\pm0.11$    &    $-5.53$    &    ?    &    ?    &    [18-20] \\  
18  &    J2030+4415    &  3.6  &    ...    &  0.07  &    $30.49\pm0.18$    &    $-3.97$    &    Y    &    N    &    [27,28] \\ 
19  &    J1741--2054    &  1.1     &    $\sim75^{\circ}$    &  0.5  &    $30.21\pm0.02$    &    $-3.76$    &    Y    &    ?    &    [29,30] \\  
20  &    J2124--3358    &  0.8   &    ...    &  0.04  &    $28.98\pm0.15$    &    $-4.85$    &    Y    &    Y    &   [21,26]  \\ 
21  &    J0357+3205    &  1.3      &    $>70^\circ$    &  1.3  &    $30.07\pm0.20$    &    $-3.70$    &    N    &    N    &    [31,32]  \\  
22  &    J0437--4715    &  0.28      &    $\sim58^\circ$    &  ...  &    $\sim$28.6    &    $-6.2$    &    Y    &    ?    &    [25,26]  \\ 
23  &    J2055+2539$^{\rm b}$    &    $<1.4$      &    ...    &    $\sim1.7$    &    $30.17\pm0.03$    &    $-3.53$    &    ?    &    ?    &    [49]  \\ 
24  &    B1929+10    &  1     &    $\sim60^\circ$    &  1.5  &    $29.50\pm0.25$    &    $-4.09$    &    ?    &    Y    &    [22,23]   \\ 
25  &    B2224+65    &    $\sim0$        &    ...    &  0.6  &    $30.18\pm0.10$    &    $-2.89$    &    Y    &    N    &     [24]  \\  
26  &    IC443$^{\rm b}$    &    $\sim 15$        &    ...    &  0.65  &    $32.82\pm0.03$    &    ...    &    N    &    Y    &    [41-43]  \\ 
27  &    MSH 15--56    &  29      &    ...    &  3.5  &    $32.8\pm0.2$    &    ...    &    ?    &    Y    &    [44,45]  \\  
28  &    G327.1--1.1    &  99      &    ...    &  5.6  &    $33.09\pm0.10$    &    ...    &    ?    &    Y    &    [46-48]  \\  
\end{tabular}
\caption{Estimated parameters of SPWNe: 
 bow shock apex stand-off distance $r_{\rm BS}$, the inclination angle $i$ between the pulsar spin axis and our line of sight, the  tail length $l$, the X-ray tail luminosity $L_X$ (in the 0.5--8 keV band), and the corresponding X-ray efficiency $\eta_X=L_X/\dot{E}$.
References:
[1] -- \citet{Wang01},
[2] -- \citet{Moon04},
[3] -- \citet{Zeiger08},
[4] -- \citet{Romani05},
[5] -- \citet{Gaensler04},
[6] --  \citet{Hales09},
[7] -- \citet{Yusef-Zadeh05},
[8] --   Klingler {\sl et al.} (in prep),
[9] -- \citet{Ng10},               
[10] -- \citet{Kargaltsev08b},
[11] -- \citet{Klingler16a},
[12] -- \citet{Petre02},
[13] -- \citet{Frail96},
[14] -- \citet{Chatterjee09},
[15] -- \citet{Ng07b},
[16] -- \citet{McGowan06},
[17] -- \citet{Klingler16b},
[18] -- \citet{Caraveo03},
[19] -- \citet{Pavlov10},
[20] -- \citet{Posselt17},
[21] -- \citet{Hui06},
[22] -- \citet{Becker06},
[23] -- \citet{Misanovic08},
[24] -- \citet{Hui12},
[25] -- \citet{Deller08},
[26] -- \citet{Brownsberger14},
[27] -- \citet{Camilo09},
[28] -- \citet{Marelli15},
[29] -- \citet{Romani10},
[30] -- \citealt{Auchettl15},
[31] -- \citet{DeLuca13},
[32] -- \citet{DeLuca11},
[33] -- \citet{Stappers03},
[34] -- \cite{Huang12},
[35] -- \citet{Kaspi01}, 
[36] -- \citet{Blazek06}, 
[37] -- \citet{Halpern14}, 
[38] -- \citet{Tomsick12},
[39] -- \citet{Pavan14},
[40] -- \citet{Pavan16}, 
[41] -- \citet{Gaensler06},
[42] -- \citet{Acciari09},
[43] -- \citet{Swartz15},
[44] -- \citet{Plucinsky02},
[45] -- \citet{Temim13},
[46] -- \citet{Slane04},
[47] -- \citet{Acero11},
[48] -- \citet{Ma16},
[49] -- \citet{Marelli16},
[50] -- \citet{Gaensler98},
[51] -- \citet{KramerJohnston08},
[52] -- \citet{Voisin16},
[53] -- this work, 
[54] -- \citet{Manchester05}.
 {\bf Note --} $^{\rm a}$ Radio emission is seen ahead of the pulsar, possibly due to interactions with the SNR shell; no radio tail is seen.  The SNR which contains the PWN is seen in H$_{\rm \alpha}$, making it challenging to discern whether the H$_{\rm \alpha}$ emission is produced by SNR reverse shocks or the pulsar wind [2]. $^{\rm b}$ The supersonic nature of the PWNe powered by these pulsars has not been firmly established.  
}
\label{tab:kd2}
\end{center}
\end{table}

\section{Magnetic Fields and Related Parameters }
PWN magnetic fields can be estimated from the synchrotron  
luminosity and the spectral slope measured in a chosen emitting region.
For a power-law (PL) spectrum with a photon index $\Gamma$, the magnetic field is given by the equation\footnote{This equation turns into equation 7.14 in \citet{Pacholczyk70} at $\nu_m = \nu_1$,  $\nu_M = \nu_2$, and $\sigma_s=3/4$, which minimizes the total energy density $w_B + w_e$.  Such equations are applicable at $\Gamma > 2/3$, the minimum value of the photon index for optically thin synchrotron radiation for any electron spectral energy distribution \citep{Ginzburg64}.}

\begin{equation}
B=\left[\frac{L(\nu_m,\nu_M)\sigma_s}{\mathcal{A} V} \frac{\Gamma-2}{\Gamma-1.5}\frac{\nu_1^{1.5-\Gamma}-\nu_2^{1.5-\Gamma}}{\nu_m^{2-\Gamma}-\nu_M^{2-\Gamma}} \right]^{2/7}\,.
\label{synch_magn_field}
\end{equation}
Here $L(\nu_m , \nu_M)$ is the synchrotron luminosity measured in the frequency range $\nu_m < \nu < \nu_M$ from a radiating volume V,
$\sigma_s = w_B / w_e$ is the ratio of the energy density of  magnetic field to that of  electrons,
$\mathcal{A} = (e^2/3m_ec^2)^2 (2\pi e m_ec)^{-1/2} = 3.06\times 10^{-14}$ in c.g.s units, 
 and $\nu_1$ and $\nu_2$ are the characteristic synchrotron frequencies ($\nu_{\rm syn} \simeq eB\gamma^2 / 2\pi m_e c$) which correspond to the boundary energies ($\gamma_1 m_e c^2$ and $\gamma_2 m_e c^2$) of the electron spectrum ($dN_e / d\gamma \propto \gamma^{-p} \propto \gamma^{-2\Gamma+1}$; $\gamma_1 < \gamma < \gamma_2$).
 The magnetic fields in SPWNe 
 estimated from synchrotron brightness
and spectral slope measurements with the aid of  
equation  \ref{synch_magn_field}
 are 
 in the range of $\sim (5$--$200)\sigma_s^{2/7}\ \mu{\rm G}$ 
(see 
 \citealt{Auchettl15}; \citealt{Klingler16a}; \citealt{Klingler16b}, \citealt{Pavan16}). 
 An additional uncertainty in these estimates is caused, in many cases,
by unknown values of the boundary frequencies $\nu_1$ and $\nu_2$, between which
the spectrum can be described by a power law.
 For $\Gamma > 1.5$ (as observed in most of the tails), the estimate of $B$ is
not sensitive to the $\nu_2$ value as long as 
$(\nu_1/\nu_2)^{\Gamma-1.5}\ll 1$.
 Varying 
  $\nu_1$ from radio ($\sim1$ GHz) to X-ray frequencies  changes $B$  by a factor of  2--3, for 
 $1.5 < \Gamma < 2$.

The magnetic field strengths can be used to 
 estimate the Lorentz factors $\gamma$, gyration radii $r_g$, and characteristic cooling times $t_{\rm syn}$ 
 of electrons radiating photons of energy $E$:
 
\begin{equation}
\gamma \sim 10^8 (E/1\ {\rm keV})^{1/2} (B/20\ \mu{\rm G})^{-1/2}\,,
\end{equation}
\begin{equation}
r_g \sim 10^{14} (E/1\ {\rm keV})^{1/2} (B/20\ \mu{\rm G})^{-3/2}\ {\rm cm}
\end{equation}

\begin{equation}
t_{\rm syn} \sim 400 (E/1\ {\rm keV})^{-1/2} (B/20\ \mu{\rm G})^{-3/2}\ {\rm yr}\,.
\end{equation}

\section{SPWN Heads: Connection to Viewing Angle and Pulsar Magnetosphere Geometry}

\begin{figure}
\centerline{\includegraphics[scale=0.46]{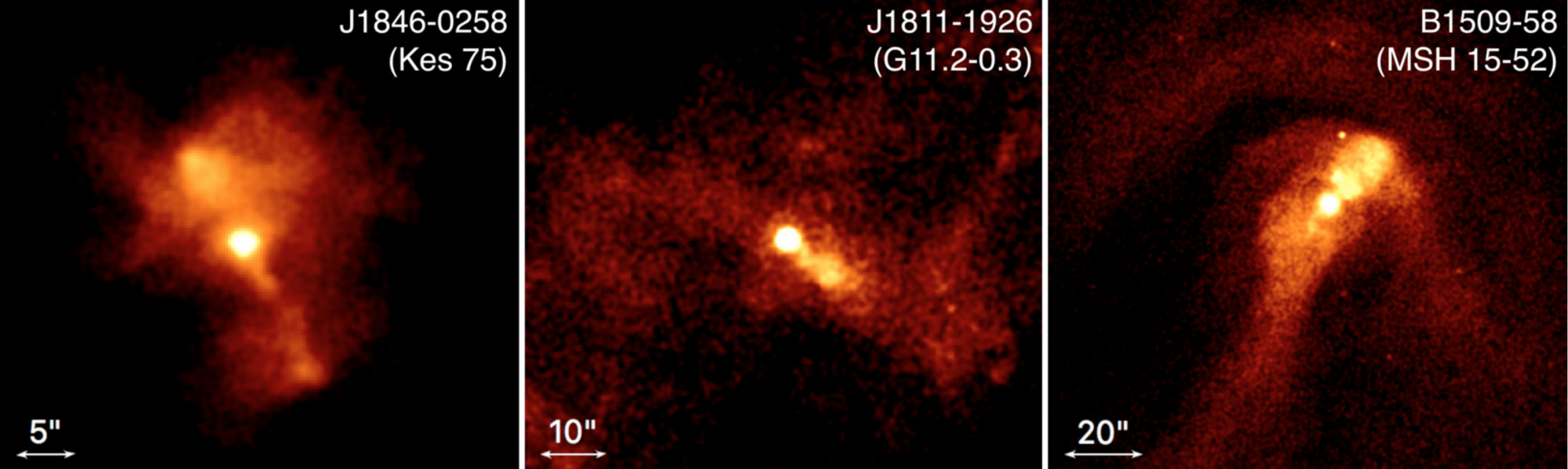}}
\caption{{\sl Chandra} ACIS images of PWNe where axial  outflows (along the pulsar spin axis) dominate equatorial components (tori).
}
\label{fig:jets}
\end{figure}

As the initially strongly magnetized wind flows away from the pulsar's
magnetosphere, its magnetic energy is likely to be at least partly converted into the 
 particle energy. 
According to the currently-popular 
 models, 
this
occurs due to magnetic field reconnection in
 a region around the equatorial plane. 
 If the pulsar is an oblique rotator (i.e., its magnetic dipole axis
is inclined to the spin axis), one expects  ``corrugated''  current sheets to be formed,  
 with regions of oppositely directed magnetic fields susceptible to reconnection (see e.g., \citealt{Lyubarsky01} for details). 
 Since the size of the reconnection region  
is expected to be larger for larger obliquity angles $\alpha$,  
the magnetic-to-kinetic energy conversion may be more efficient
for pulsars with larger $\alpha$.
 Such pulsars may exhibit brighter PWNe, 
 with more pronounced equatorial components 
(e.g., the Crab PWN).  
It is less clear what 
 happens at small $\alpha$; likely, the magnetic-to-kinetic energy conversion would still take place, but  
outflows along the spin axis
 (``jets'') would be more pronounced than the equatorial
``tori''\footnote{The coexistence of a strong polar outflow with a comparatively weak equatorial one can be reproduced in the \citet{Komissarov04} model, 
 in which the jet is formed by backflow of the equatorial wind  
diverted by the magnetic field hoop stress toward the spin axis.}
 \citep{2016MNRAS.462.2762B}.  
Examples of such  morphologies for young, subsonic pulsars
  are the  G11.2--0.3, Kes 75, and MSH 15--52 PWNe, which    exhibit relatively  
less luminous tori in the X-ray images (see Figure \ref{fig:jets}).  
For SPWNe, the identification of the equatorial and polar components is more challenging.

In the X-ray images shown in Figure \ref{fig:bow:xray} one can often (but not always; see examples below) identify relatively bright and compact PWN ``heads'' accompanied by  much dimmer (in terms of surface brightness) extended tails.  Recent deep, high-resolution {\sl Chandra} observations 
 revealed  fine structures of several bright  heads  
with contrasting morphologies (cf.\ the insets in panels 10 and 16 in Figure \ref{fig:bow:xray}).  
For instance, the head of the B0355+54 PWN  looks like a symmetric, filled ``dome'',
 brighter near the axis than on the sides \citep{Klingler16b}. 
 In contrast, the head of the J1509--5058 PWN, 
looks like two bent tails \citep{Klingler16a}, with almost no emission between them, except for
a short southwest extension just behind the pulsar\footnote{
 Interestingly, the tails look slightly separated from
 the pulsar similar to
 the base of  the south-eastern inner jet  in the Vela PWN;
 see Figure 1 in \cite{Levenfish13}.}.
 This structure is remarkably similar to 
the Geminga PWN (\citealt{Posselt17}; see panel 17 in Figure \ref{fig:bow:xray}),
just the angular size of the latter is larger, in accordance with the smaller
distance (250 pc vs.\ 4 kpc). 
The bow-shaped X-ray emission can 
 be associated with either
a limb-brightened 
 shell 
 formed by the shocked PW downstream of the TS
  or pulsar jets bent by the ram pressure of the oncoming ISM. 
In the former case, a lack of diffuse emission in between the
lateral tails 
 would require 
 a nonuniform magnetic field in the emitting shell, possibly caused by  amplification of the ISM magnetic field component perpendicular to the pulsar's velocity vector
 \citep{Posselt17}.
In the latter scenario, the winds of J1509--5058 and Geminga must be dominated by luminous polar components (as in the PWNe shown in Figure \ref{fig:jets})
   rather than by the equatorial component (as in the Crab and Vela PWNe). 
  If the 
 lateral tails of the J1509--5850 head are indeed bent jets, it may be difficult to explain the ordered helical magnetic field morphology in the extended tail (as suggested by radio polarimetry; \citealt{Ng10});   
such a structure would be more natural 
 for the axially symmetric case \citep{Romanova05}, when the pulsar spin axis
(hence the jet directions) is co-aligned with the velocity vector. 

\begin{figure}
\centerline{\includegraphics[scale=0.52, angle=0]{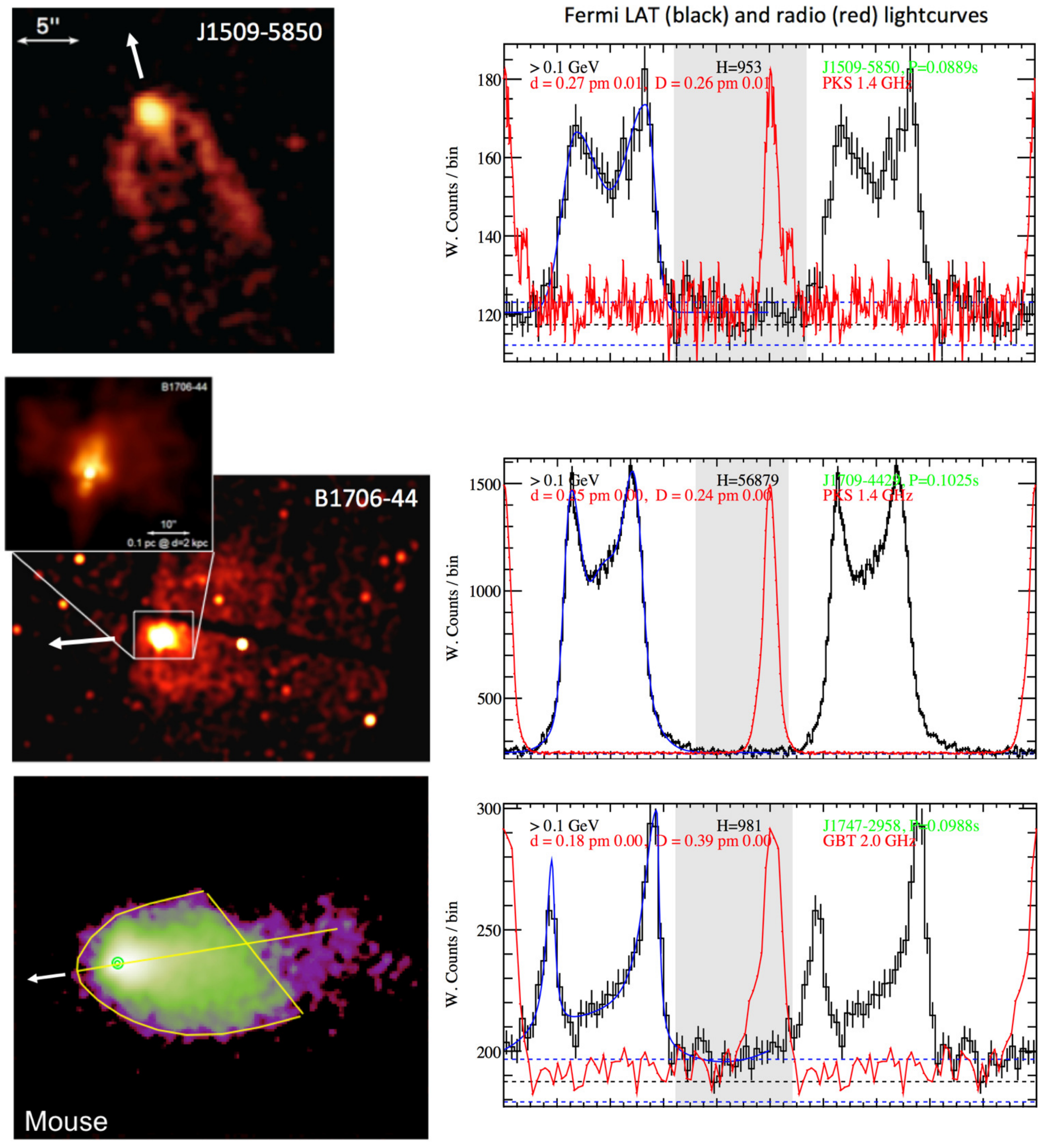}}
\caption{Comparison of the  PWN morphologies and pulsar light curves for PSRs J1509--5850,  B1706--44, and  J1747--2858 (top to bottom). The very similar light
 curves of all three  pulsars (both in radio and $\gamma$-rays) suggest similar 
angles between the spin axis and the line of sight, $\zeta$,
 and between the spin and magnetic axes, $\alpha$. 
 The contour drawn on top of  the Mouse PWN image  represents a possible extent of the equatorial outflow affected by the ram pressure (the outflow is in the plane of the shown contour which is symmetric with respect the pulsar velocity direction but appears to be asymmetric once projected onto the sky; Klingler et al.\ in prep.). }
 \label{fig:b1706}
\end{figure}

The quite different ``filled''  morphology of the B0355+53 PWN head could be due to different mutual orientations of the pulsar's spin axis, magnetic dipole axis, velocity vector, 
and the line of sight. 
Some information on these orientations could be inferred from the pulsar's signatures at different
energies. 
For instance, Geminga is a $\gamma$-ray pulsar (as well as PSR B1509--58) with no bright radio emission, while PSR B0355+64 was not detected in $\gamma$-rays but is quite bright in radio. 
One can speculate that Geminga and J1509--5850 are moving in the plane of the sky, and their spin axes are nearly perpendicular to the velocity vectors and to the line of sight
(this assumption would be consistent with the jet interpretation of the lateral tails). 
On the contrary, the spin axis of B0355+54 could be nearly aligned with the line of sight, in which case the ``dome'' would be interpreted as the equatorial torus distorted by the ram pressure while the central brightening would be the sky projection of the bent jets.  
Thus, it is quite plausible that the qualitative morphological differences in the appearances of PWN heads can be attributed to geometrical factors (i.e., the angles between the line of sight, velocity vector, spin axis, and magnetic dipole axis).

Since the pulsar light curves in different energy ranges should also depend on the same
 geometrical factors, it is interesting to look for correlations
between the PWN head shapes and light curves. For instance,
both the $\gamma$-ray and radio light curves of PSRs J1509--5850 and B1706--44
are remarkably similar, not only in shapes but also in phase shifts between
the $\gamma$-ray and radio pulses
(see Figure \ref{fig:b1706}),
 which implies similar geometries and allows one to expect similar PWN morphologies.
 We see in Figure \ref{fig:b1706} that the B1706--44
X-ray PWN shows clear jets (without obvious bending in the pulsar vicinity)
 and a relatively underluminous equatorial component 
 (in contrast to the Crab). 
Although faint, the large-scale  morphology of the B1706--44 PWN 
 suggests that the pulsar is moving at the position angle (east of north) of about $80^{\circ}$ \citep{Ng08}. 
Thus, although the PWN head morphologies do not look exactly the same,
one could imagine that if B1706--44 were moving faster and had even more pronounced jets, 
 its PWN in the pulsar vicinity  could look like the one around J1509--5850.  
The radio and $\gamma$-ray lightcurves of the  Mouse pulsar (J1747--2958) are similar to those of PSRs J1509--5058 and J1709--4429 (see Figure \ref{fig:b1706}).  The only difference is that the $\gamma$-ray pulse of PSR J1747--2958 is slightly wider (with a deeper trough) and more asymmetric compared to those of PSRs J1509--5058 and J1709--4429.  
All three pulsars display single radio peaks with very similar phase separation from the $\gamma$ pulses.   
According to the outer gap magnetospheric emission models, this implies a fairly large magnetic inclination angle
(so that both $\gamma$-ray and radio pulsations can be seen).  
It is likely that these angles are somewhat larger for PSR J1747--2958 than those of PSRs J1509--5058 and J1709--4429.  
These considerations help to interpret the appearances of the compact nebulae by suggesting that the equatorial outflow dominates over the polar outflow components in these cases.

Interestingly, we do not see bright heads in the X-ray PWNe of some supersonic pulsars, even those with relatively bright X-ray tails (e.g., PSR J1101--6101 and J0357+3205; panels 9 and 21 in Figure \ref{fig:bow:xray}). 
In some of those cases, however, the heads may be too compact to resolve because of high pulsar velocities and large distances.

\begin{figure}
\centerline{\includegraphics[scale=0.42]{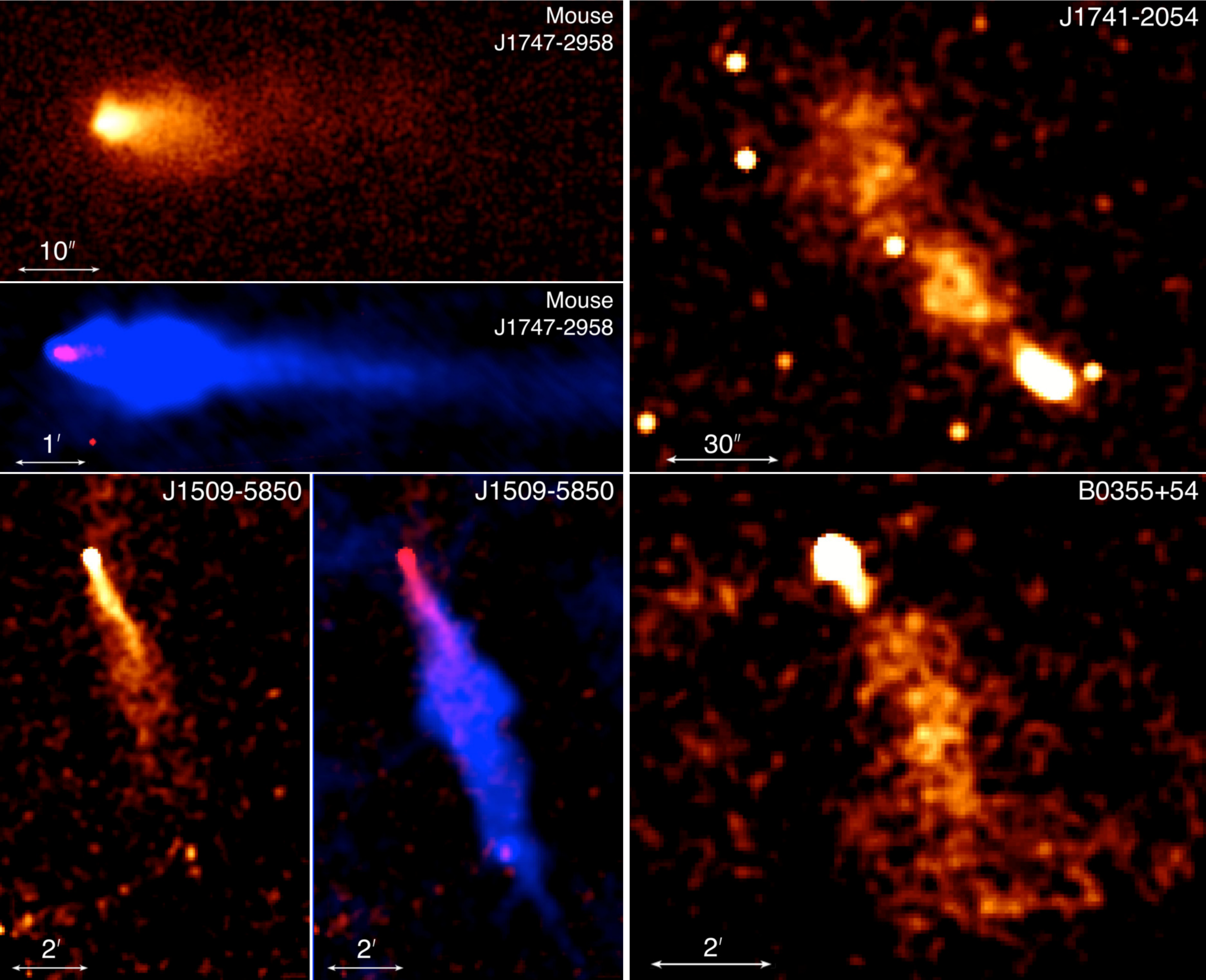}}
\caption{Extended tails behind four pulsars. The X-ray images are obtained with {\sl Chandra} ACIS.  For the Mouse PWN and the J1509--5850 PWN, combined X-ray (red) and radio (blue) images are shown.}
\label{fig:tails}
\end{figure}

 SPWNe can also display contrasting radio morphologies. 
X-ray bright PWN heads may or may not be bright in radio. For instance the  Mouse PWN 
  has the  head which is  bright in both X-rays and radio,
 while there is very little (if any) radio emission from the head  of the J1509--5058  PWN, although it is bright in X-rays (see Figure \ref{fig:tails}).

\begin{figure}
\centerline{\includegraphics[scale=0.27]{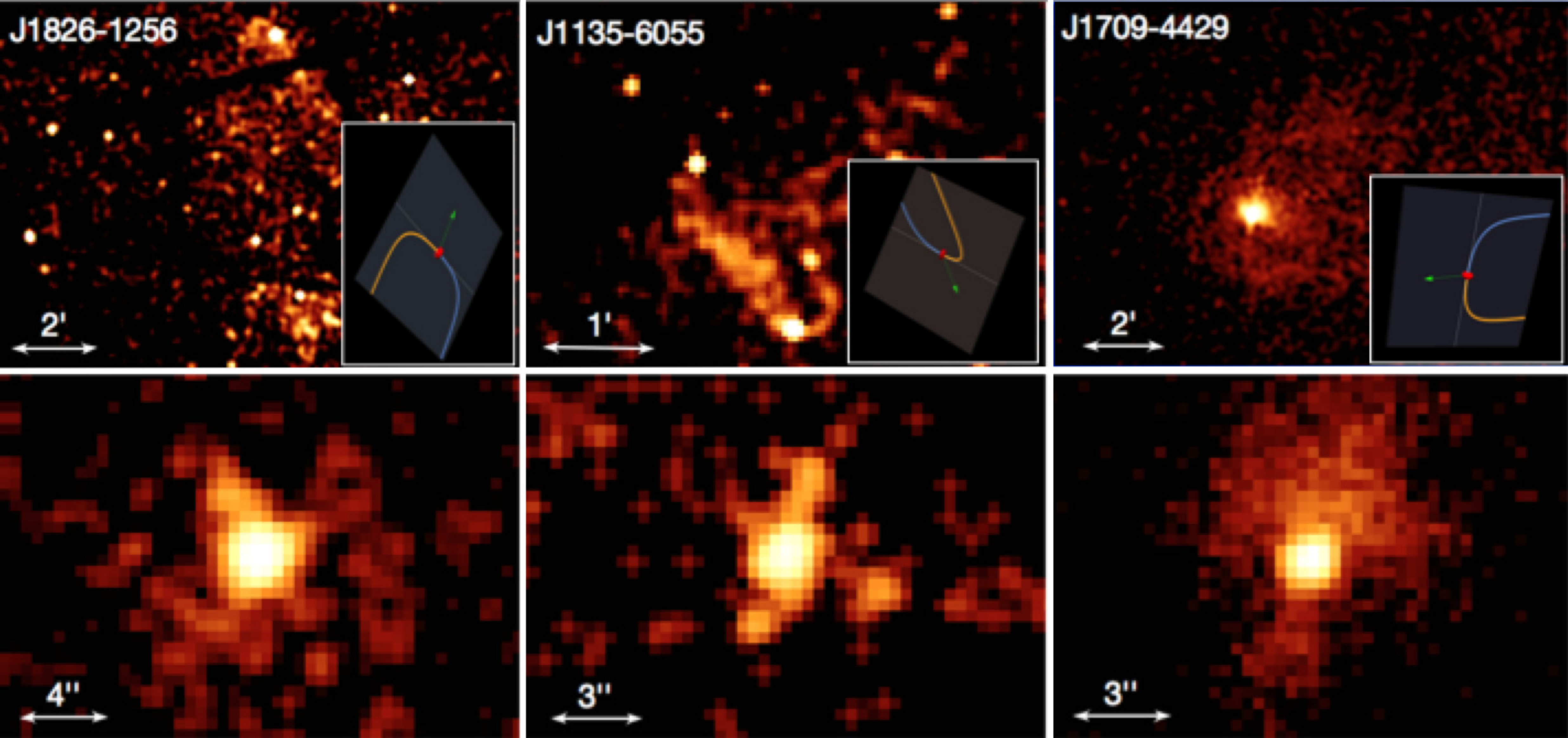}} 
\caption{\chan\ images of  SPWNe that likely move with mildly supersonic velocities. A schematic diagram of a possible geometry is shown for each object, with the jets bent by the ram pressure.  The green arrows indicate the inferred direction the velocity vector.}
\label{fig:models}
\end{figure}

Understanding the causes for the differing head (and, possibly, tail; see \S5) morphologies in SPWNe is important because it can help to determine the orientation of the pulsar spin axis with respect to the line of sight and to the pulsar velocity vector (see Figure \ref{fig:models}). 
The former is the angle $\zeta$, an important parameter for   comparing magnetospheric emission models with the $\gamma$-ray and radio light curves  (see e.g., \citealt{Pierbattista16}). 
The angle between the spin axis and  the pulsar velocity (the direction of the ``natal'' kick for not too old pulsars) has important implications for the supernova explosion models  \citep{Ng07}. SPWNe are particularly suitable for this purpose because the projected NS velocity can be inferred from the PWN morphology in the absence of NS proper motion measurement. 
 
\section{Pulsar  Tails}
\begin{figure}
\centerline{\includegraphics[scale=0.6]{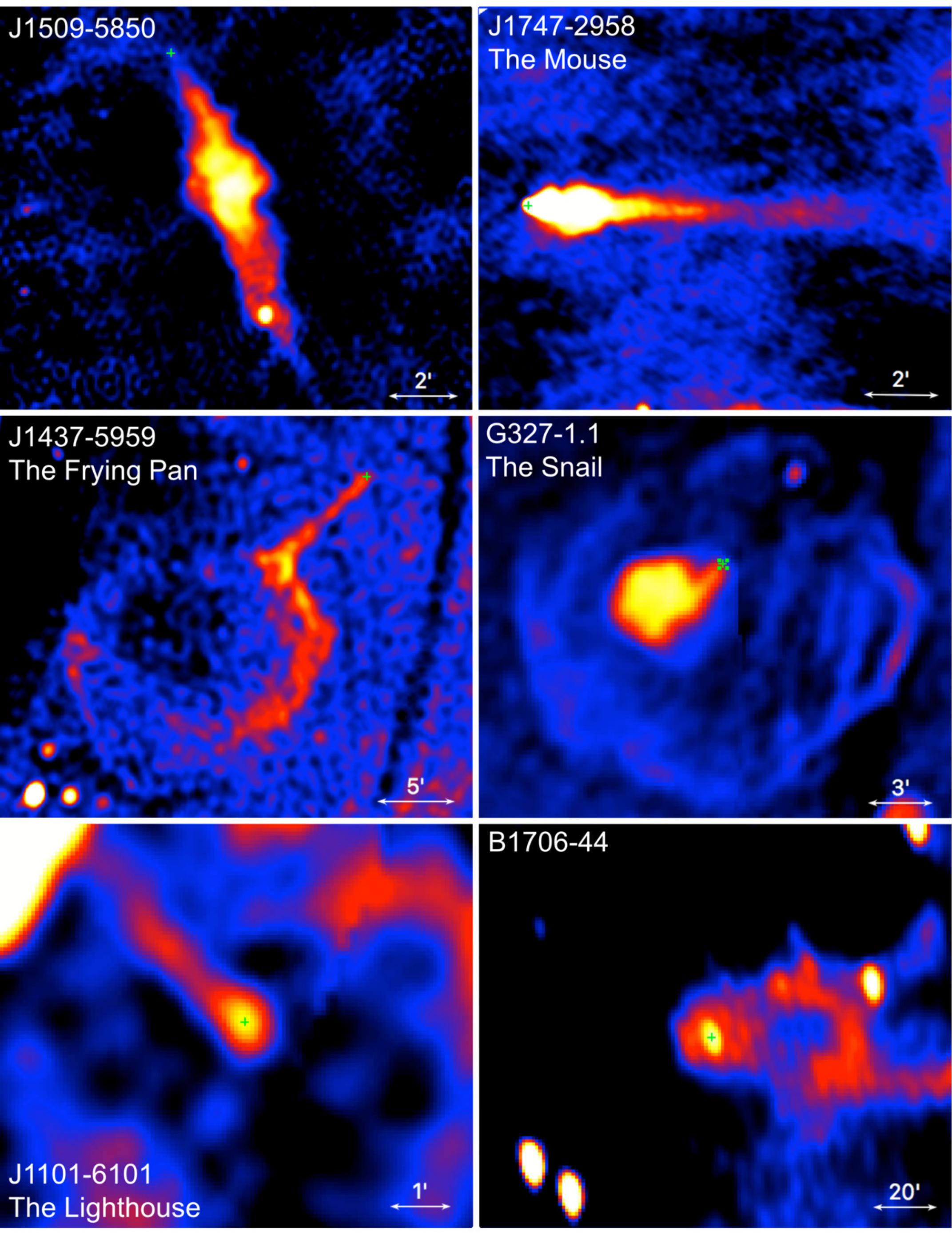}}
\caption{Radio images of pulsar tails: J1509--5850 (ATCA, 5 GHz), J1747--2958 (VLA, 1.5 GHz), J1437--5959 (MOST, 843 MHz), G327--1.1 (MOST, 843 MHz), J1101--6101 (MOST, 843 MHz), and B1706--44 (VLA, 1.4 GHz).  The green crosses mark the positions of the pulsars (for the Snail no pulsations are detected and the cross shows the position of the X-ray point source).}
\label{fig:radio_tails}
\end{figure}

As supersonic pulsars move through the ISM, the ram pressure confines and channels the PW in the direction opposite to the pulsar's relative velocity with respect to the local ISM.  
Therefore, on large spatial scales (compared to the TS bullet size) one expects to see a {\em pulsar tail} --  an extended, ram-pressure-confined structure behind the pulsar (see Figs.\ \ref{fig:tails} and \ref{fig:radio_tails}).
Several pulsar tails have now been discerned above the background for up to a few parsecs from their parent pulsars. 
The longest known tails are those of PSR J1509--5850, whose projected visible length spans 7 pc (at $d=3.8$ kpc) in X-rays and $\sim 10$ pc in radio (\citealt{Klingler16a}, \citealt{Ng10}), and the Mouse PWN, whose X-ray and radio tails span 1 and 17 pc, respectively (at $d=5$ kpc; \citealt{Gaensler04}). 

On large scales,  the shapes of many pulsar tails (e.g., B1929+10 -- \citealt{Wang93}, \citealt{Becker06}, \citealt{Misanovic08}; J1509--5850 -- \citealt{Klingler16a}; J1437--5959   -- \citealt{Ng12}) can be crudely approximated by cones that widen with distance from the pulsar as the outflow slows down and expands (see Figure \ref{fig:bow:xray}).
However, in some cases the tail brightness can be strongly nonuniform showing more complex structures that can be described as expanding ``bubbles'' (e.g., the SPWNe of PSR J1741--2054; see panel 19 in Figure 2;  also \citealt{Auchettl15}. 
The bubbles and the rapid widening of tails can be attributed to nonuniformities in the ISM, instabilities in the  backflow from the pulsar bow shock \citep{vanKerk08}, or entrainment (mass loading of the ISM into the PW; \citealt{Morlino15}).
Some tails, such as those of PSRs J1741--2054  \citep{Auchettl15}  and B0355+54 \citep{Klingler16b}, exhibit noticeable ``bendings'' at large distances from the pulsars, which could be attributed to ISM winds.  
Although {\sl Chandra}, with its superior angular resolution and low background, delivers better images than {\sl XMM-Newton} for most pulsar tails, some tails have been studied with {\sl XMM-Newton} as well (e.g., B1929+10 -- \citealt{Becker06};  J2055+2539 -- \citealt{Marelli16}).

Radio polarimetry of two extended tails (the Mouse and J1509--5058; \citealt{Yusef-Zadeh05} and \citealt{Ng10}) shows that the magnetic field direction is predominantly transverse in  the J1509--5058 tail while it is aligned with the tail in the case of the Mouse. 
This could indicate that the spin axis is more aligned with the velocity vector in J1509--5058 than in J1747--2858 (see Figure 3 in \citealt{Romanova05}). 
This, however, would be at odds with the jet interpretation of the lateral outflows in the J1509--5058 PWN head.

Furthermore, the brightness of the radio tail in the Mouse decreases with distance from the pulsar, whereas in the J1509--5850 and J1101--6101 tails the radio brightness increases with the distance from the pulsar and peaks around 4 pc and 1.7 pc (at $d=3.8$ kpc for J1509--5850 and $d=7$ kpc for J1101--6101).  
Such radio surface brightness behavior could be explained if the PWN magnetic field becomes stronger further down the tail. It is possible for a helical magnetic field configuration if the flow velocity, $v_{\rm tail}$, decreases rapidly enough with distance from the pulsar, which may lead to an increase in the magnetic field strength ($B_{\rm tail}\propto v_{\rm tail}^{-1}S_{\rm tail}^{-1/2}$, where $S_{\rm tail}$ is the cross-sectional area of the tail; see, e.g., \citealt{Bucciantini05}).  
The G327.1--1.1 (Snail) PWNe contains an undetected pulsar whose tail radio brightness remains constant over its $1\farcm5$ length until it connects to a spherical structure (most likely a reverse shock, as the PWN is located inside its supernova remnant; see Figure \ref{fig:radio_tails}; \citealt{Ma16}).

The PWN of PSR J1437--5959 (the Frying Pan; Figure \ref{fig:radio_tails}) is not seen in X-rays, although it is prominent in radio and displays a long radio tail which fades with distance from the pulsar until it becomes brighter near the shell of the alleged SNR G315.9--0.0 (see  \citealt{Ng12}). 
Another possible example of such a PWN is the one of  PSR  B0906--49 (J0908--4913) which has been discovered in radio  \citep{Gaensler98} but was not detected  in the subsequent {\sl Chandra} ACIS observation \citep{KargaltsevChaps}. It is currently unclear what makes these PWNe so underluminous in X-rays. 
A natural explanation could be a lack of sufficiently energetic particles which might be due to a magnetosphere geometry unfavorable for accelerating particles to high energies. 
However, for PSR B0906--49, radio timing observations suggest that it is an orthogonal rotator with the pulsar's spin axis and direction of motion aligned \citep{KramerJohnston08}, thus making it similar to some pulsars with bright X-ray PWNe.

Another puzzling property is the apparent faintness of pulsar tails in TeV 
\footnote{TeV emission has been recently detected from Geminga, but it is likely produced in a large-scale ``halo'' rather than in the tail; see \citealt{Abeysekara17} and \citealt{Linden17})} 
while IC TeV emission has been detected from many younger, more compact PWNe (see e.g., \citealt{Kargaltsev13} for a review).   
In the leptonic scenario (IC up-scattering of Cosmic Microwave Background radiation, dust IR, and starlight photons off relativistic electrons and positrons), the PWN TeV luminosity should not depend on particle density of the surrounding medium. 
The lack of detections can be attributed to the limitations of the current TeV observatories that have poor angular resolution and  may not be sensitive to narrow long structures such as pulsar tails\footnote{ 
The sizes of  circular extraction regions typically used  in TeV data analysis substantially exceed the transverse sizes of the tails due to the coarse resolution of current TeV  imaging arrays.
 Under these circumstances the faint TeV emission from the tail could be buried in the background.}. 

\begin{figure}
\centerline{\includegraphics[scale=0.42]{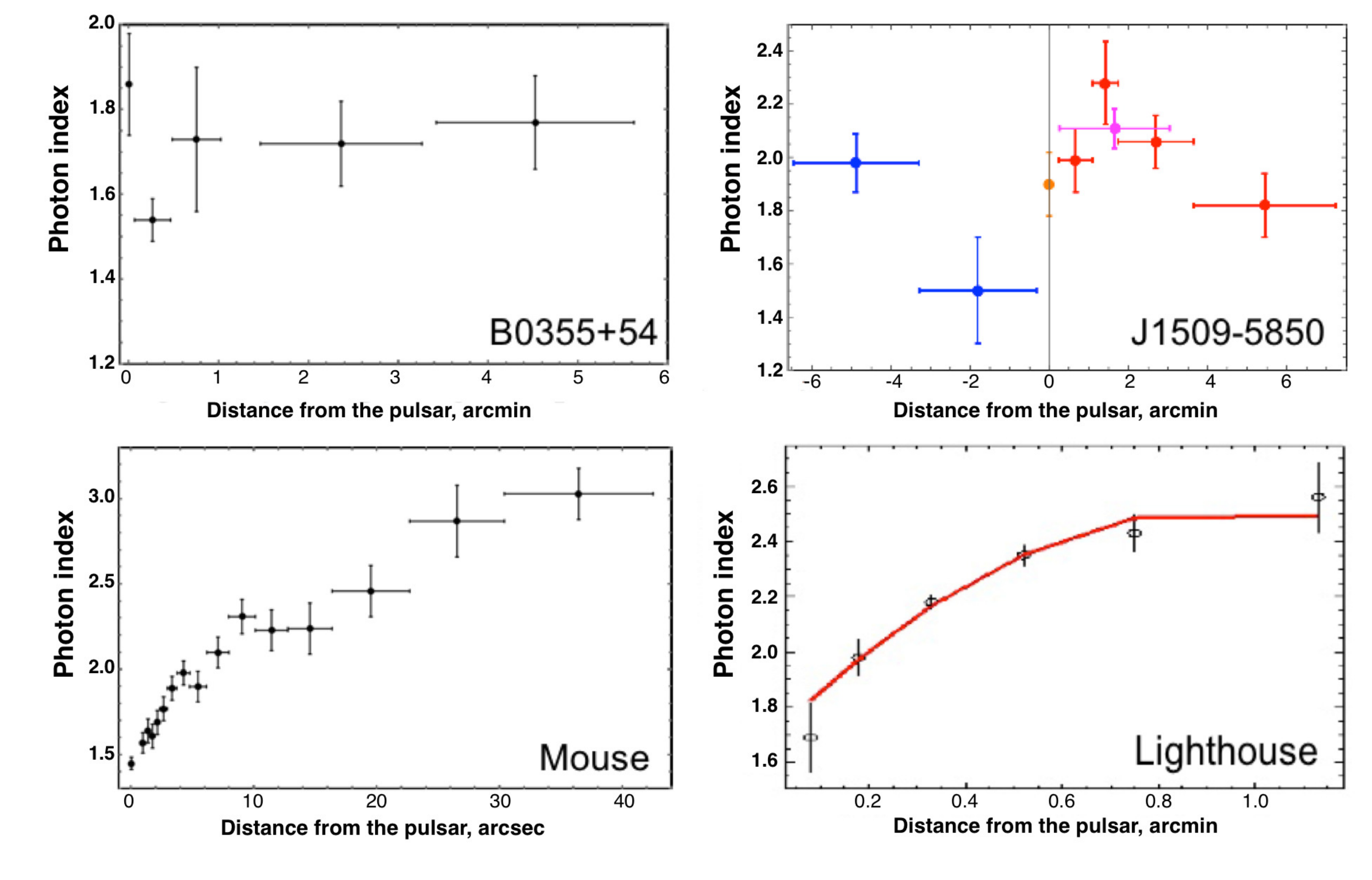}}
\caption{Variation of spectral slopes along pulsar tails. No cooling trends are seen in B0355+54 and J1509--5850, while cooling (spectral softening) is very pronounced
in the Mouse (Klingler et al., in prep.) and Lighthouse \citep{Pavan16}, perhaps due to higher magnetic fields, lack of in-situ acceleration, or slower flow speed. The red line in the Lighthouse panel shows the best fit with a  parabolic function (from  \citealt{Pavan16}).}
\label{fig:spectral}
\end{figure}

Yet another puzzle of pulsar tails is the very different dependences of their X-ray spectra on the distance from the pulsar along the tail (see Figure \ref{fig:spectral}). 
The rapid changes of  photon index, likely due to synchrotron cooling, are seen in the tails of younger pulsars (e.g., J1747--2958, J1101--6101, and J0537--6910) while virtually no changes are seen in the tails of older pulsars (e.g., J1509--5850 and B0355+54). 
The different spectral evolution might be explained by different strength of magnetic field. 
The magnetic field strengths could be reduced by more efficient reconnection in the tail in the case of pulsars with a large angle between the spin axis and the velocity vector because larger angles may lead to a more tangled large-scale magnetic field in the tail. 
The continuing reconnection of magnetic fields in long tails could also lead to particle re-acceleration, which could help to explain the lack of softening in the X-ray spectra of the J1509--5850 and B0355+54 tails. 
In cases where the spin axis is nearly aligned with the magnetic dipole axis, reconnection may be delayed (because magnetic energy is not being efficiently converted to particle energy) until significant distortions or turbulence develop at larger distances down the tail (which may explain the lack of a ``head' in the X-ray images of J0357+3205). 
A larger sample of bow shock PWNe is needed to probe these scenarios.

From Equation (3.4) in \S3, it is  possible to crudely estimate the average characteristic flow speed  in the tail: 
$v_{\rm flow} \sim l / t_{\rm syn}$, where $l$ is the length of the X-ray tail.
The bulk flow speeds estimated  in this simplistic way  are 2000, 3000, 3000, and 20000 km s$^{-1}$ in the B0355+54,  J1509--5850, J1101--6101, and the Mouse tails, respectively\footnote{An additional uncertainty in these estimates comes from the unknown inclination angle of the velocity with respect to the line of sight.  For these estimates we used the values that were considered plausible in the corresponding individual publications (see Table 2). }. 
 They are significantly higher that the pulsar speeds (with a possible exception of the  very fast PSR J1101--6101).

\section{Misaligned Outflows} 

\begin{figure}
\centerline{\includegraphics[scale=0.3]{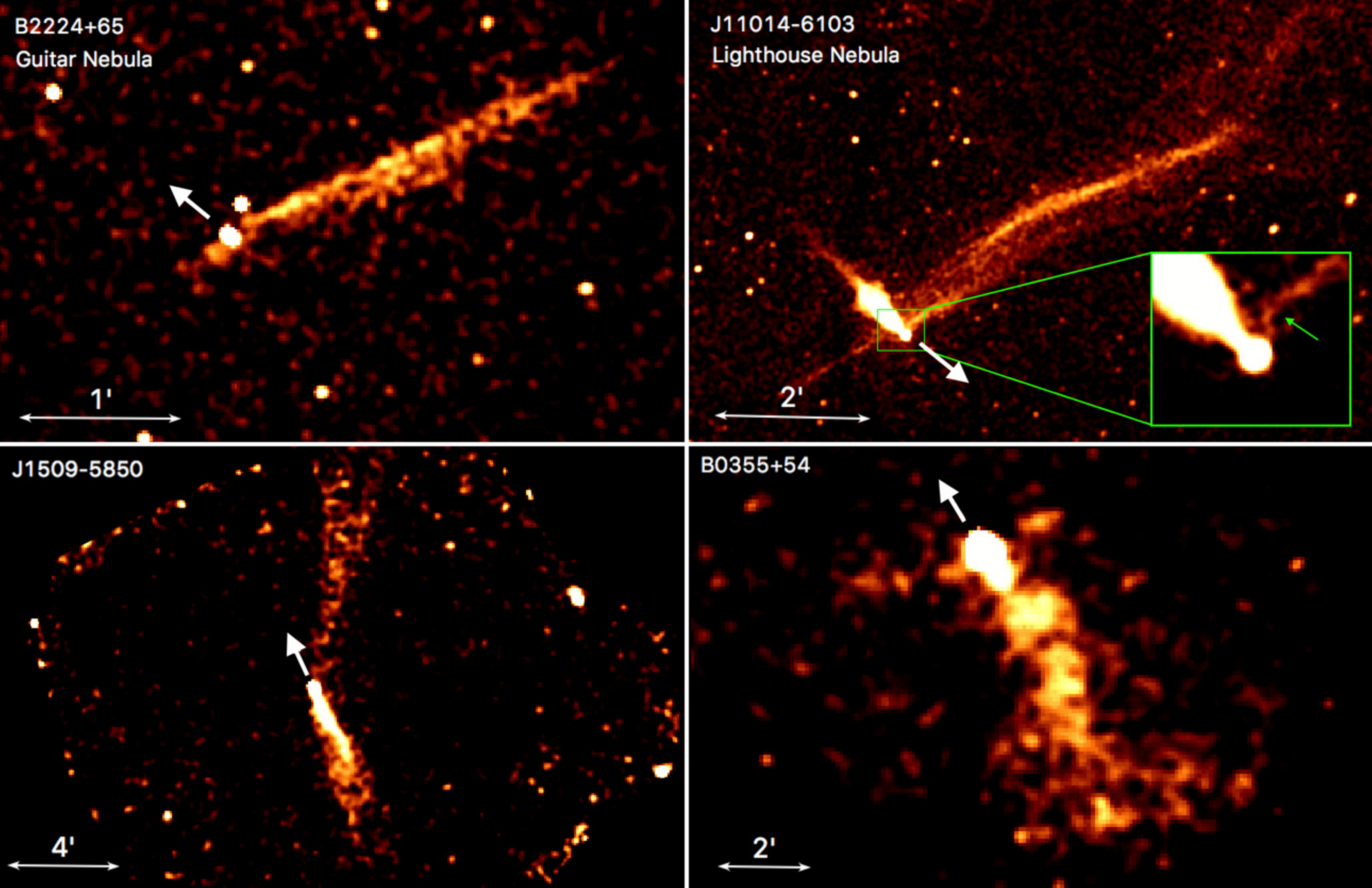}}
\caption{{\sl Chandra} images of PWNe displaying misaligned outflows.  The white arrows 
 show the directions of pulsar proper motion, and the green arrow shows the bending in the Lighthouse Nebula outflow (inset). {\sl Chandra} images of some of these  objects are also shown in  \citet{Reynolds17}.}
\label{fig:misaligned_outflows}
\end{figure}

In recent deep X-ray observations, a new type of structure has unexpectedly been discovered in some supersonic PWNe. 
Extended, elongated features, {\em strongly misaligned with the pulsar's direction of motion}, are seen originating from the vicinity of four pulsars (see Figure \ref{fig:misaligned_outflows}):  the Guitar Nebula  (PSR B2224+65; \citealt{Hui07}), the Lighthouse PWN (PSR J1101--6101; \citealt{Pavan14}, 2016), the PWN of PSR J1509--5058 \citep{Klingler16a}, and the Mushroom PWN of PSR B0355+54 (Klingler et al.~2016b). 
Another, possibly similar, misaligned feature was reported for PSR J2055+2539 based on {\sl XMM-Newton} observations \citep{Marelli2016}. 
The misaligned orientation of these features is puzzling because, for a fast-moving pulsar, one would expect all of the PW to be confined within the tail (which three of the four PWNe exhibit as well). 
In principle, one could imagine a highly anisotropic wind with extremely strong polar outflows (jets) misaligned with the pulsar's velocity vector. 
Such jets should be bent by the ram pressure of the oncoming ISM, with the bending length-scale $l_b\sim\xi_j\dot{E}/(c r_j \rho v^2 \sin\theta)$, where $\xi_j<1$ is the fraction of the spin-down power $\dot{E}$ that goes into the jet, $r_j$ is the jet radius, and $\theta$  is the angle between the initial jet direction and the pulsar's velocity. 
However, the bending length-scales are way too small compared to the lengths of the nearly straight misaligned features.
For instance, $l_b\sim 0.065 \xi_j (r_j/5.7\times10^{16} {\rm cm})^{-1} n^{-1} (\sin \theta)^{-1} (v/300\,{\rm km\,s}^{-1})^{-2}$~pc for J1509--5850, while the observed length of the misaligned outflow is $\sim$8 pc (for $d=3.8$ kpc; see \citealt{Klingler16a} for details).
Therefore, it is difficult to explain the misaligned outflows with a hydrodynamic-type model.

To interpret the first discovered misaligned outflow in the Guitar Nebula,
 \citet{Bandiera08} 
 suggested that such structures can be formed in high-$\mathcal{M}$ pulsars
 when the gyro-radii of most energetic electrons, $r_g=\gamma m_e c^2/eB_{\rm apex}$, are comparable to or exceed the stand-off distance of the bow-shock apex,
 $R_{\rm h} = (\dot{E} /4\pi c  m_p n v^2)^{1/2}$ (for isotropic wind), 
 where   $B_{\rm apex}$ is the magnetic field inside the PWN
(between the CD and TS) 
  near the bowshock apex. 
 Such particles cannot be contained within the bow shock, and can hence ``leak'' into the ISM, where they diffuse along the ambient ISM magnetic field lines and radiate synchrotron photons.   The Lorentz factors of  escaping electrons, therefore,  can be estimated as  
$\gamma\sim 2\times10^{8}(E/1\,{\rm keV})^{1/2}(B_{\rm ISM}/5\,\mu{\rm G})^{-1/2}$, where $B_{\rm ISM}$ is the ambient magnetic field, and $E$ is  the synchrotron photon energy that reaches at least 8 keV for 
 the Lighthouse, Guitar, and J1509--5058 PWNe. From the escape condition, $r_g \gtrsim R_{\rm h} $,  one can set an upper limit on the PWN field near the apex:
 $B_{\rm apex}\lesssim34(E/1\,{\rm keV})^{1/2}(B_{\rm ISM}/5\,\mu{\rm G})^{-1/2}(R_{\rm h}/10^{16}~{\rm cm})^{-1}$~$\mu$G, if $R_{\rm h}$ is directly measured or estimated from the equation given above.

It is important  to stress that in 
 this scenario  the misaligned outflows are an entirely kinetic phenomenon,
 which makes them  very different from the pressure-confined jets of tails. The observed appearances of the  misaligned outflows should reflect the ambient ISM structure\footnote{If the leaking particles carry sufficiently large currents, the currents can perturb the original ambient ISM magnetic field. } 
 illuminated by synchrotron emission from the leaked  PW particles. In the discussed scenario, the misaligned structures are expected to move with the particle injection site (SPWN apex) which moves with the pulsar.

The Bandiera (2008) scenario 
was challenged by recent deep {\sl Chandra} observations of the Lighthouse PWN where the misaligned outflow appears to bend around the 
 pulsar (as indicated by the green arrow in the inset image in the top right panel of Figure \ref{fig:misaligned_outflows}). 
Although the distortion of the external (ISM) field can be expected if a magnetized object is moving through  the ISM (``magnetic draping''; \citealt{Lyutikov06} 
and  \citealt{Dursi08}), this should not be happening outside the forward shock region where the ISM ``does not know''  about the incoming object yet. 
 Another difficulty with this scenario is that most of these misaligned outflows (with a possible exception of a very faint one associated with B0355+54) are strongly asymmetric (i.e., they appear much brighter and longer on one side of the pulsar).

\section{Forward Shock Emission}
If the local ISM contains neutral Hydrogen, collisional excitation and charge exchange in the forward shock can cause the shocked ISM to emit in H$\alpha$ leading to formation of an H$\alpha$ bow shock.
To date, bow-shaped H$\alpha$ nebulae have been detected around only 8 rotation-powered pulsars (see Table 2 and \citealt{Brownsberger14}).
 A possible explanation for the scarcity of H$\alpha$ bow shocks is a high degree of ionization of the ambient medium ahead of the pulsar, which might be caused by pre-ionization of the ISM by radiation from the pulsar and/or the X-ray PWN. 

\begin{figure}
\centering
\includegraphics[scale=0.68]{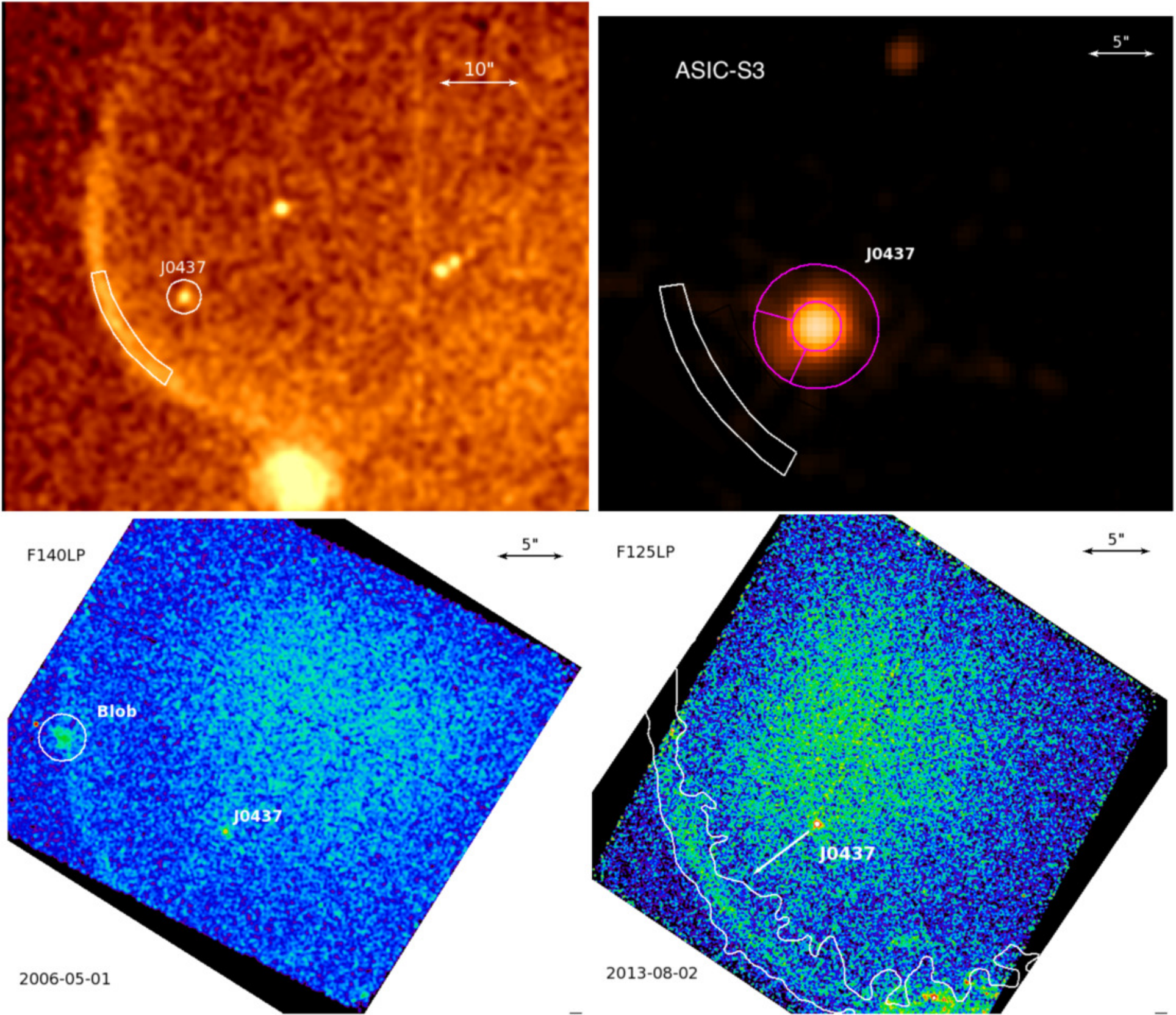}
\caption{The top left panel shows an H$\alpha$ image of PSR J0437--4715  
 obtained with the SOAR telescope \citep{Brownsberger14}.
 The top right panel shows a {\sl Chandra} ACIS-S3 image. 
 The magenta contours delineate the source and background extraction regions used for the extended emission analysis; the white contour, coinciding with that
in the top left panel, shows the position of the H$\alpha$ bow shock apex.
 The two bottom  panels show FUV images 
 from two different observations (dates are labeled) and  different filters (also labeled).  The nonuniform background prominent in 
 the FUV images is due to the ``thermal glow'' of the detector.  The extended ``blob'' of unknown nature is 
 seen in the bottom left panel.  H$\alpha$ bow shock contours (based on the SOAR image) are shown in the  bottom right panel.  North is up, East is to the left. The images are taken from \citet{Rangelov16}.}
\label{j0437}
\end{figure}

\begin{figure}
\includegraphics[scale=0.64]{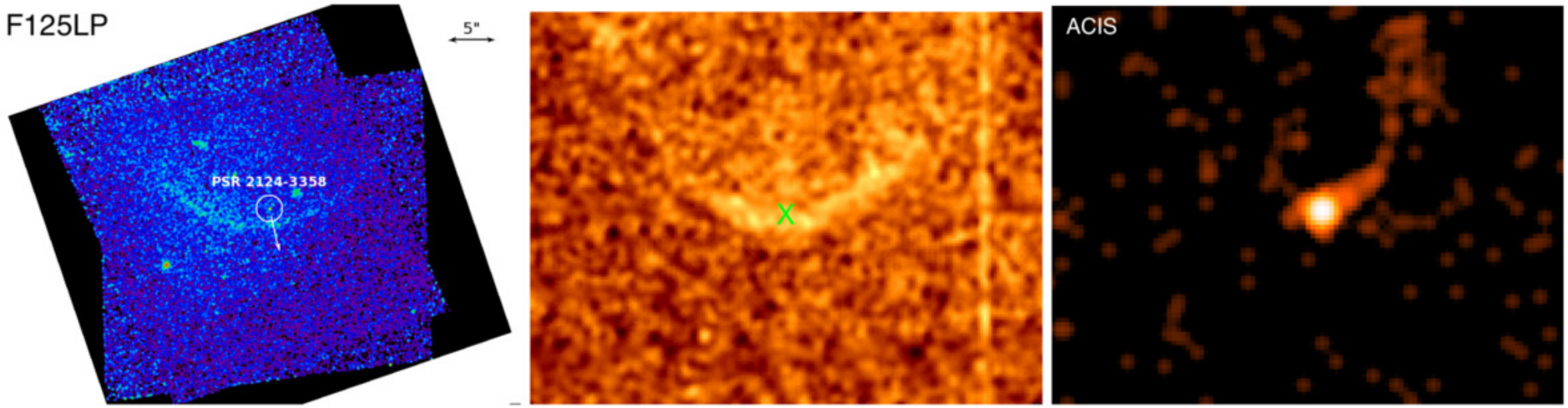}
\caption{The left panel shows a far-UV image of the PSR J2124--3358 vicinity obtained with the {\sl Hubble Space Telescope (HST)} ACS/SBC F125LP filter. 
The pulsar and its direction of motion are shown with white circle and arrow. 
The middle panel shows the SOAR H$\alpha$ image; ``X'' marks the position of the pulsar.
The right panel shows the {\sl Chandra} ACIS image in 0.5$-$7\,keV. 
All images are to the same scale.  North is up, and East to the left. The images are taken from \citet{Rangelov17}.}
\label{j2124}
\end{figure}

One of the first H$\alpha$ pulsar bow shocks was detected by \citet{Bell93} 
 ahead of the binary millisecond (recycled) pulsar J0437--4715.
 PSR J0437--4715 is the closest known pulsar, $d=156.8\pm0.2$\,pc \citep{Reardon16}, with an accurately measured proper motion
  corresponding to the transverse velocity $v_\perp=104.1\pm0.2$\,km\,s$^{-1}$ \citep{Deller08}. It has a period of 5.8 ms and a spindown power
 $\dot{E}=2.9\times10^{33} I_{45}$\,erg\,s$^{-1}$ (corrected for the Shklovskii effect; $I_{45}$ is the neutron star moment of inertia in units of $10^{45}$ g cm$^2$).
 Its H$\alpha$ bow shock shows a symmetric structure with a standoff distance of about $10''$ ($2.3\times 10^{16}$\,cm)
 between the 
 pulsar and the  
bow shock apex (see Figure \ref{j0437}).

The J0437--4715 bow shock is the first one detected in far-UV (FUV), thanks to the pulsar proximity and deep observations 
 with the {\sl Hubble Space Telescope (HST)}.
 The FUV images 
 reveal a bow-like structure positionally coincident with the 
  H$\alpha$ bow shock (Figure~\ref{j0437}). 
 The  measurements in broad F125LP and F140LP filters (wavelength bands are 1250--2000 \AA\ and 1350--2000 \AA, respectively) do not allow us to establish the spectral model. The existing data are consistent with  both 
  a simple power-law spectrum, which could be continuum synchrotron emission from relativistic electrons trapped in the FS region \citep{Bykov17}, 
  and the   collisionless plasma emission (spectral lines plus continuum) from the ISM matter compressed and heated in the FS  region \citep{Bykov13}. 

Analyzing a subset of archival  {\sl Chandra} ACIS data, \citet{Rangelov16} also found a faint extended X-ray emission ahead of the pulsar but well within the H$\alpha$ bow shock (Figure \ref{j0437}). 
This emission was tentatively interpreted as an X-ray PWN of J0437. 
The morphology of the PWN is unusual compared to other 
 supersonic PWNe in the sense that the X-ray emission 
 is seen only ahead of the moving pulsar, being
 confined to a relatively narrow region. 
However, such a bright region of emission ahead of the pulsar 
 consistent with simulated synchrotron maps of the bow shock PWNe (see Figure 4 from \citealt{Bucciantini05}).  
These regions may simply be unresolvable in other, more remote PWNe, but due to the proximity of J0437,
 {\sl Chandra} observations can probe shocked PW X-ray emission on much smaller spatial scales.

The second bow shock detected in FUV 
 is associated with  the solitary millisecond  PSR J2124--3358 
 \citep{Rangelov17}). 
 This pulsar has a period of 4.93\,ms and  a spin-down 
 power $\dot{E}=6.8\times10^{33}I_{45}$\,erg\,s$^{-1}$.
 It is at a distance $d=410^{+90}_{-70}$\,pc, 
 its proper motion corresponds to $v_\perp= 101.2\pm 0.8$ km, at $d=410$ pc
 \citep{Reardon16}. 
The shape of its FUV bow shock 
 matches well the H$\alpha$ bow shock, which suggests that the FUV emission comes from the forward shock region. 
The poor statistics
 once again prevents us from 
 discriminating between the PL and collisionless shock plasma emission models. 
A sensitive slit or integral field spectroscopy would enable further progress in understanding the nature of FUV emission from the pulsar bow shocks.

\citet{Hui06} reported 
 an X-ray PWN associated with J2124--3358. 
 It was detected with both {\sl XMM-Newton} and {\sl Chandra}, but only {\sl Chandra} was able to resolve the nebula into a one-sided 
 narrow structure, which appears to be projected inside the observed H$\alpha$-FUV bow shock (see Figure~\ref{j2124}).  
The X-ray emission extends northwest from the pulsar for $\sim0\farcm5$. 
\citet{Hui06} modeled the spectrum of the emission with a PL model with photon index $\Gamma=2.2\pm0.4$.

In addition to 8 rotation-powered pulsars,
bow-shaped H$\alpha$ emission has been 
 detected around the radio-quiet isolated neutron star RX~J1856.5--3754
\citep{vanK01}.
 It belongs to the class of neutron stars  
 that do not show any signatures of  magnetospheric activity (e.g.,
nonthermal X-ray emission or
 radio/$\gamma$-ray pulsations; see \citealt{Kaplan11} and references therein).
 In this case the H$\alpha$ emission  may represent a photoionization nebula 
 rather than a bow shock \citep{vanK01}.  For this reason we do not include RX~J1856.5--3754 in Tables 1 and 2.

\section{Conclusions}

SPWNe exhibit a number of interesting properties providing insight into the structures of neutron star magnetospheres, winds, and pulsar viewing geometry, as well as into the physics of collisionless shocks and supernova explosions.
Some of these  properties became apparent  only from deep {\sl Chandra } and {\sl HST} observations:

\begin{itemize}
\item Very long (up to 8 pc) X-ray tails have been detected behind a few pulsars.  In some cases the X-ray  tails are accompanied by even longer radio tails.  None of these long tails are seen in TeV.
\item Some of the tails exhibit very rapid spectral softening (cooling) with increasing distance from the pulsar (e.g., the Mouse and Lighthouse PWNe), while others show no evidence of cooling (e.g, J1509--5850). 
\item When resolved in X-rays, the SPWN head morphologies  can differ drastically. There can be ``filled'', ``empty'', and very dim (of absent) SPWN heads. The differences can be linked to the geometry of the pulsar magnetosphere and to the orientation of pulsar's spin axis with respect to the observer's line of sight.  
\item There is a growing number of SPWNe with misaligned outflows which are likely produced by most energetic particles leaking from the SPWN apex. However, there are some unsolved puzzles in this scenario. 
\item Far-UV emission from two H$\alpha$ bow shocks has been recently discovered, likely produced by heated shocked ISM.
\end{itemize}

\smallskip
Although a general,  qualitative  picture of PWNe created by supersonically
moving pulsars is more or less clear, there remain a number of open questions.

\begin{itemize}
\item Why are some SPWN heads very faint in X-rays?  Are they simply too compact to be resolved from the pulsar?
\item Why do X-ray spectra of some tails remain hard at large distances from the pulsar?
\item What are particle acceleration mechanisms in SPWNe, and is there in-situ acceleration outside the TS?
\item How some of the tails remain so well collimated for such long distances?
\item What is the nature of the misaligned outflows? Why are they  so asymmetric?  
\item Are the morphologies of X-ray PWN heads and PWN radiative efficiencies and spectra uniquely determined by mutual orientations of the line of sight, pulsar's velocity, spin, and magnetic axis?
\item Are the SPWNe intrinsically fainter in TeV $\gamma$-rays compared to PWNe inside the SNRs?
\end{itemize}

\medskip
To answer these and other questions,  detailed studies of a larger sample of SPWNe must be  supplemented by  more realistic  models.

\section{Acknowledgments}

This work was partly supported  by the NASA through
Chandra Awards   GO3-14057 and G03-14082  issued  by  the
Chandra X-ray  Observatory Center, which is operated by the Smithsonian
Astrophysical Observatory for and on behalf of the NASA under contract
NAS8-03060.  It was also supported by grants GO-12917 and GO-13783 from the Space Telescope Science Institute, 
which is operated by Association of Universities  for Research in Astronomy, Inc.\  under NASA contract    NAS  5-26555.
 We thank Igor Volkov for discussions and help with the figures.

\bibliographystyle{jpp}

\bibliography{references}

\end{document}